\documentclass[aip,reprint,onecolumn]{revtex4-1}
\usepackage{amsmath, amssymb}
\usepackage[normalem]{ulem}
\usepackage{etex}
\usepackage{graphicx}
\usepackage[T1]{fontenc}
\usepackage[utf8x]{inputenc}
\usepackage{lmodern}
\usepackage{color}

\begin{document}
\title{Heat transfer enhancement on thin wires in superfluid helium forced flows}
\author{Davide Dur\`{i}}
\affiliation{Univ. Grenoble Alpes, LEGI, F-38041 Grenoble, France}
\affiliation{CNRS, LEGI, F-38041 Grenoble, France}
\affiliation{Univ. Grenoble Alpes, INAC-SBT, F-38000 Grenoble, France}
\affiliation{CEA, INAC-SBT, F-38000 Grenoble, France}
\author{Christophe Baudet}
\affiliation{Univ. Grenoble Alpes, LEGI, F-38041 Grenoble, France}
\affiliation{CNRS, LEGI, F-38041 Grenoble, France}
\author{Jean-Paul Moro}
\affiliation{CEA, DEN-DANS-DM2S-STMF-LIEFT, F-38000 Grenoble, France}
\author{Philippe-Emmanuel Roche}
\affiliation{Univ. Grenoble Alpes, Institut NEEL, F-38042 Grenoble, France}
\affiliation{CNRS, Institut NEEL, F-38042 Grenoble, France}
\author{Pantxo Diribarne}
\email{pantxo.diribarne@cea.fr}
\affiliation{Univ. Grenoble Alpes, INAC-SBT, F-38000 Grenoble, France}
\affiliation{CEA, INAC-SBT, F-38000 Grenoble, France}


\begin{abstract}
  In this paper, we report the first evidence of an enhancement of
  the heat transfer from a heated wire by an external turbulent flow of
  superfluid helium. We used a standard Pt-Rh hot-wire anemometer and
  overheat it up to 21 K in a 
  pressurized liquid helium turbulent round jet at temperatures
  between 1.9 K and 2.12 K. The null-velocity response of
  the sensor can be satisfactorily modeled by the counter flow
  mechanism while the extra cooling produced by the forced convection
  is found to scale similarly as the corresponding extra cooling in
  classical fluids. We 
  propose a preliminary analysis of the response of 
  the sensor and show that -contrary to a common assumption- such
  sensor can be used to probe local velocity in turbulent superfluid helium. 
\end{abstract}

\maketitle


\section{Introduction}
At temperatures below $T_\lambda \approx 2.18\,\mathrm{K}$ liquid
helium undergoes a phase transition, from a classical fluid
(He~I) to quantum one (He~II). The latter phase exhibits many peculiar
properties among which the ability to flow without apparent
dissipation through thin capillaries, the quantization of the
vorticity in atomic-diameter vortex filaments,  and a very efficient heat
transport associated with the existence of temperature waves (``second
sound'')\cite{VanSciverLivre2012}.  
Those properties are well understood in the framework of the
two-fluid model: He~II is described as a mixture of a ``normal fluid'' and a ``superfluid''.
The first one is viscous and carries all the entropy of the fluid, while the second is inviscid and irrotational except on the quantum vortices. 

The present work was motivated by the experimental study of
turbulence\cite{Vinen02, Barenghi14Intro} in He~II, and associated
velocity sensors. In classical fluids, a common picture of turbulent
motion is the Richardson cascade: the largest 
eddies  of the flow are continuously stretched producing 
smaller and smaller eddies with no significant energy loss in the ``cascade'' process. This
process holds until eddies are small enough for the viscous
dissipation to become significant. A good illustration of this
process is given by the distribution of the kinetic energy
as a function of the flow scale. In He~II, at large scale,
experimental, theoretical, 
and numerical works (e.g. see ref. \cite{BarenghiPNAS14} for a review)
indicates that the two components of He~II are locked (through the
so-called "mutual friction" term), and exhibit Kolmogorov-like power spectra. 
Unfortunately, the smallest scales of turbulent flows are far from
being resolved by the existing velocity sensors.

To the best of our knowledge, only two major velocity measurement
principles were applied to He~II: (\textit{i}) stagnation pressure
measurements, such as total head pressure tubes~\cite{Maurer98,Salort10}, Pitot
tubes~\cite{Salort10,Salort12} or cantilevers~\cite{Salort12b};
(\textit{ii}) particle
visualization methods among which Laser Doppler Velocimetry~\cite{Murakami89},
Particle Image Velocimetry\cite{Zhang04} and Particle Tracking
Velocimetry~\cite{Bewley06, *LaMantia13}. 

One of the most commonly used velocity sensors in classical turbulence is
absent from the above list: the hot-wire. The 
popularity of hot-wires in the field of hydrodynamics is due to their very high
spatial and temporal resolution~(see e.g. \cite{Chanal00,Bailey10} for extreme
miniaturization). Such sensors have already proven very useful at cryogenic
temperatures~(e.g. \cite{Moisy:1999p3738,Chanal00,Pietropinto:2003p331,QuixQuest:HotWireETW:2012}),
in particular for the study of very high Reynolds number
turbulence in gaseous helium. However, it is often assumed that
hot-wires cannot work in He~II~\cite{Vinen10}. Their measurement
principle is based on the heat transfer enhancement when a heated wire
is submitted to forced convection. In classical fluids, at length
scales typical of hydrodynamic experiments, the forced convection
is much more efficient than natural convection and molecular
diffusion. However, in superfluid helium, another very efficient
thermal transfer mechanism comes to be even more
efficient than forced convection. Physically, this mechanism is associated
with the generation of a counter-flow between the normal and
superfluid components of He~II (respectively subscripted $n$ and $s$
hereafter). The normal component flows outward, carrying its entropy
away from the heat source, while the entropy-less superfluid component
flows inward and ensures net mass conservation. The efficiency of this
heat transfer 
mechanism is limited by the generation of a superfluid vortex tangle
self-sustained by the counter-flow itself. Nevertheless, even
for large heat fluxes, where the vortex line density is large, 
the counter-flow mechanism stands very
efficient and forced convection does not improve significantly the heat
transfer\cite{Johnson:1978}. For instance, hybrid magnets can be 
cooled by static He~II~\cite[p.~54]{Barenghi01}, even with long
distances between the thermal source and sink.  

The need for time and space resolved velocity measurements
has led us to analyze to which extent the heat transfer improvement due
to forced convection in He~II can be resolved. A standard Ag coated Pt-Rh
Wollaston wire driven by a commercial constant temperature anemometer
was used to investigate the effect of forced convection on
heat transfer in He~II. In this article, we show that the response of
the sensor can be successfully used as a local velocity measurement. 
After a brief description of the experimental apparatus, we first analyze
the static response of the sensor, i.e. without external flow. The
standard approach for describing heat transfers in He~II, is found to
satisfactorily predict the response of the sensor. In the presence of
an external flow we observe an enhancement of the heat transfer. We
present 3 experimental evidences that the local velocity of the external
flow is responsible for the enhancement.

\section{Measurement setup}
\subsection{Wind tunnel}
The experiment is performed in a pressurized cylindrical vessel
($\varnothing 200 \,\text{mm}\times 500 \,\text{mm}$) where a liquid
helium round jet develops from a nozzle with inside diameter
$D_n=5\,\text{mm}$ (see~\cite{Duri11} for details). The temperature
can be continuously varied from 4.2~K to 1.7~K, so that classical
(He~I) and superfluid (He~II) flows can be achieved in the same
apparatus respectively above and below the superfluid transition 
temperature $T_\lambda \simeq 2.17 K$. The pressure is kept greater than 
the helium critical pressure ($P_C \approx 2.2\,\text{bar}$) in order 
to avoid the onset of boiling at the surface of the hot-wire. 

The hot wire is located on the axis of the jet, at $45\times D_n$ or
$60\times D_n$ downstream from the nozzle. In this region, classical
turbulence literature (see e.g.~\cite{Wygnanski69}) show that most
of the quantities of interest such as the velocity and its first
moments, the integral length scale and the Taylor's length scale, are
self-similar. 
As we could not map the velocity field of the jet we assumed classical
behavior so that the velocity at a particular distance from the nozzle
is assumed to follow the same scaling as in \citet{Wygnanski69}. This
assumption is supported by the fact that we found a
quantitative agreement between the expected integral time and the
auto-correlation time of the signal  at various Reynolds
numbers both in He~I and He~II. 

The mean velocity $U_{45D}$ or $U_{60D}$ seen by the wire is varied
typically from 0.1 up to 1.5 m/s, 
corresponding in He~I, to Reynolds number up to $Re =
U_{0D}D_n/\nu\simeq 1\times10^6$ where $\nu$ is kinematic viscosity
of He~I.

\subsection{Hot-wire}

The probes are manufactured using standard Platinum-Rhodium
(90\%Pt-10\%Rh) Wollaston wires of diameter $1.3\,\mathrm{\mu m}$. The
length of the sensitive etched part of the wires is $l_w\approx
400\,\mathrm{\mu m}$. The wire is welded on the stainless steel prongs
of a home made ceramic mounting.

From room temperature down to few tens of Kelvins, the resistivity
$\rho(T)$ of Pt-Rh alloy decreases linearly with temperature. Below 30
K, the temperature dependence roughly vanishes as $\rho(T) \simeq
\rho_0 (1 + 2 .10^{-8} T^4)$, where the residual resistivity $\rho_0$
is interpreted as being due to the presence of defects and impurities
in the material (here $\rho_0$ corresponds to $R_{wire} \approx 50
\Omega$). The hot 
wire was overheated at a temperature set within $T_{w}=20-25~$K.
In those conditions the fluid that surrounds the wire undergoes a
steep but continuous density and
temperature variation. We have checked, in He~I, that electrical
response of the hot-wire did
not show peculiar characteristics due to this very high temperature
difference with regards to the surrounding fluid temperature: 
the usual Kings calibration law holds down to the lowest velocities 
explored with He~I. 

The wire is driven at constant resistance (``Constant Temperature
Anemometry'' or CTA mode) using a commercial DISA-55-M10 CTA anemometer
bridge. The output voltage signal of the anemometer is proportional to
current required to 
overheat the hot-wire up to the setup resistance (i.e. temperature). The
resistance control system was checked to be reliable up to frequencies
of order 10~kHz so that the results presented hereafter are low-pass-filtered
at 5 kHz. 

The signals were acquired either on a NI-PXI4462 or a HP-E1430
acquisition boards, at sampling frequencies up to 
$100\,\mathrm{kHz}$. For one velocity measurement, typical data set
was 30 files with $2^{22}$ points. This corresponds to 4000 integral
times at the highest velocities.
 
\section{Results}
\label{par:res}

\subsection{DC response of the hot-wire}

\begin{figure}[!ht]
  \centering
  \includegraphics{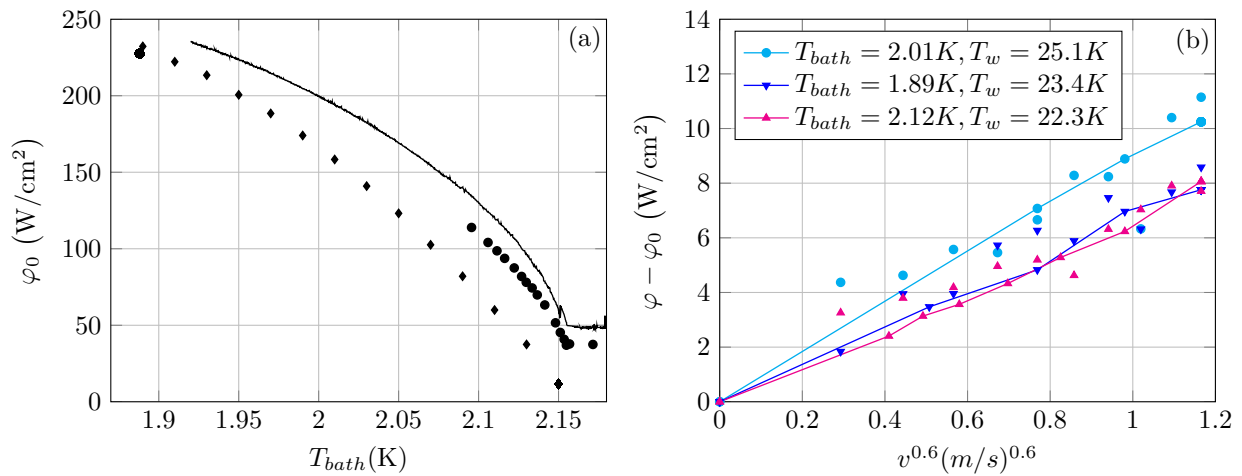}
  \caption{ (a): Average heat flux at the surface of a hot wire of size
    $D_w = 1.3 \,\mathrm{\mu m}$ operated at $T_w \simeq
    21K$ at null velocity. Continuous line:  time series during
    cooling of the bath, 
    obtained using a constant-resistance CTA electronics. Circles ($\bullet$):
    open loop constant voltage measurements (see text). Diamonds
    ($\blacklozenge$): Numerical integration of the model based on
    conduction function. (b): additional heat flux due to He~II forced
    convection, 
    represented as a function the power 0.6 of the velocity. The
    points that are linked by a line are the result of a 30 times
    longer averaging and their temperature is controlled much more
    carefully than unlinked points.}
  \label{fig:stat}
\end{figure}

The continuous line of Figure~\ref{fig:stat}(a) presents the mean heat
flux $\varphi_0$ at the surface of the hot-wire as a function of bath
temperature when the jet is turned off and the wire regulated at
$T_{w}\approx 21$ K. Right below the superfluid transition ($T_\lambda
\approx 2.15$~K at 2.6~bars), the heat flux rises sharply, with a 5
fold increase between $T_\lambda$ and 1.9~K. To discard
  possible artifact of the regulation electronics, this measurement
  has been reproduced independently in open-loop mode, by manual
  adjustment of an independent voltage 
  source driving the hot-wire (filled circles). For later discussion,
it is convenient to introduce here the bath-temperature sensitivity
$\gamma$ of the hot-wire heat flux $\varphi_0$:
\begin{equation}
\gamma=\frac{1}{\varphi_0} \frac{d\varphi_0}{dT_{bath}}
\end{equation}
Over the explored temperature range, typically $\gamma \simeq
-2\,\mathrm{K^{-1}}$. 

The observed velocity dependence of the heat flux $\varphi$ (averaged
over the whole wire surface) can be written in the same form as in
classical fluids: 
\begin{equation}
  \label{eq:calibration}
  \varphi = \varphi_0 + B v^\alpha,
\end{equation}
with $\alpha \simeq 0.5$ in the general case due to the
$v^{-0.5}$ scaling of the thermal boundary layer thickness. While in
He~I data (not shown here), $\alpha=0.5$ is well suited, in He~II
case, a value $\alpha=0.6$ gives a slightly better fit (see
figure~\ref{fig:stat}(b)). 

The probe response in He-II and its similarity with the response of
classical hot wires are first indications that hot wires can be
operated as velocity probes in He-II. Still, heat transfer processes
occurring near the wire surface are more complex -as discussed later-
and no clear justification can
be brought about the slight change in the scaling with the mean flow
velocity. Consequently, we have decided to analyze time series data of
the raw CTA output $e(t)$ rather than attempting to use Eq. 2 to
convert it into velocity time series. Since $\varphi \sim e(t)^2$ and
$\varphi -\varphi_0 \ll \varphi_0$, the fluctuations of the raw signal
$e(t)$ are nearly proportional to the heat flux $\varphi -\varphi_0$.

\subsection{Power spectra from the hot-wire}

\begin{figure}[!ht]
  \centering
  \includegraphics[width=8.5cm]{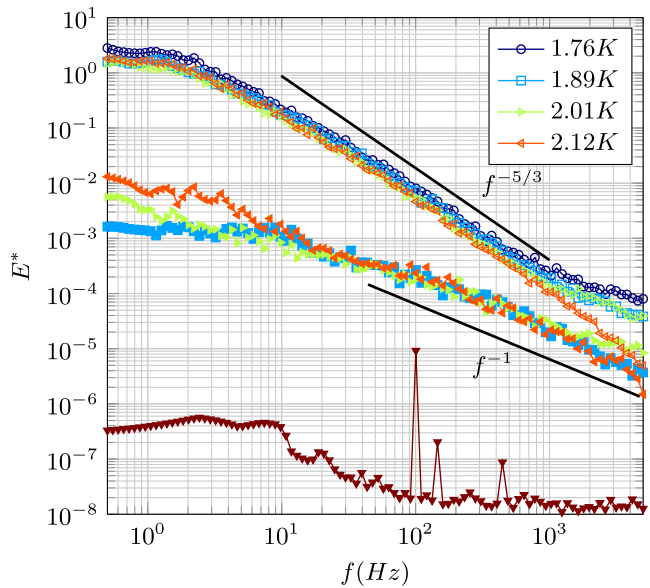}
  \caption{Power spectral density of the CTA electronics output
    voltage normalized by its variance, for a flow velocity
    $U_{60D}=1.3\,\mathrm{m.s^{-1}}$ (open symbols) at four
    temperatures. The corresponding colors with filled 
    symbols correspond to the same temperature but at null
    velocity. In the latter case, the signal is normalized by the
    variance of the signal at non-null velocity. Pointing down
    triangles show the observed noise in He~I (the corresponding non-null
    velocity spectrum is not shown here)}
  \label{fig:spectra}
\end{figure}

Figure~\ref{fig:spectra} presents the power spectral density (PSD) of
the voltage $e$ delivered by the hot wire electronics,
normalized by its variance $\sigma_e^2$:
\begin{equation}
  \label{eq:PSD}
  E^*(f) = E(f) / \sigma_e^2,
\end{equation}
where $E$, the PSD of the voltage signal, is computed using Welch
periodogram method over windows of $2^{16}$ points. The PSD at four
different temperatures, ranging from $1.76\,\mathrm{K}$ up to
$2.12\,\mathrm{K}$ ($T< T_\lambda$), are represented as a function of
the frequency for a mean flow velocity
$U_{60D}=1.3\,\mathrm{m.s^{-1}}$ on the wire. Null jet velocity PSD
are represented (filled symbols) normalized by the variance of their
corresponding non-null-velocity signal. For comparison, the normalized PSD at
null jet velocity in He~I is also represented (pointing down
triangles).

The first observation is that the null velocity ``noise'' has a much
higher  level of fluctuations 
in He~II than in He~I.
    
At non-null jet velocities, a  Gaussian white PSD is observed up to
$f\approx 2 \,\mathrm{Hz}$. This is consistent with the expected
incoherent motion of the very large scales of the flow. For higher 
frequencies, the PSD exhibit a power law behavior consistent with Kolmogorov
 $f^{-5/3}$ scaling. A departure from this power law is observed at
frequencies of order $f\approx 2\,\mathrm{kHz}$, except at
temperatures close to $T_\lambda$: the PSD decreases much less rapidly
(roughly as $ f^{-0.5}$). As can be seen, at the highest resolved
frequencies, the energy density of the signal is always higher than the one
observed at null velocity. 

The observation of a plateau followed by a Kolmogorov-like spectra in He-II
is a second indication 
that the hot wire is sensitive to velocity fluctuations, at least up
to 2 kHz. Indeed, such spectral behavior have already been reported
in He-II flows (e.g.~\cite{Maurer98,Salort10,Salort12}) and their
observation is known to be robust to non-linearities in the calibration law.
 
\begin{figure}[ht!]
  \centering
  \includegraphics[width=8.5cm]{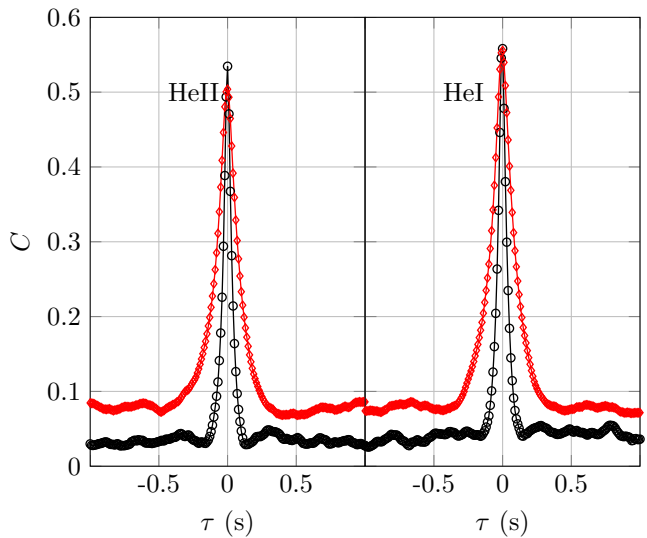}
  \caption{Cross-correlation coefficient of the hot wire signal and of
    a total head pressure tube situated 4~mm apart (black $\circ$ 1.72
    m/s and red $\diamond$ 0.86 m/s, color online). Measurements are
    performed at 45~$D_n$ from the nozzle.}
  \label{fig:correl}
\end{figure}

\subsection{Correlation with a Pitot tube anemometer}

A Pitot-tube anemometer working both in He-I and He-II was specially
made to test the correlation between the raw hot-wire signal and
velocity fluctuations. It is mounted close to the hot-wire, at a 4~mm
transverse distance and 3~mm  downstream of it. Its design is based on
those  found in~\cite{Salort10}. 
Figures~\ref{fig:correl}(a) and (b) present the cross-correlation
$C(\tau)$ of the Pitot anemometer raw electrical signal $v(t)$ with the
hot wire CTA raw output
signal $e(t)$ both in He-I and He-II, where
$$C(\tau) = \frac{\langle (e(t)-\langle e \rangle) ( v(t+\tau)
  -\langle v \rangle  ) \rangle}{\sigma_{e}\sigma_{v}},$$  
$\sigma_{e}$ and $\sigma_{v}$ being the root-mean-squared values of
the  signals. 

In He-I both probes are known to measure velocity. The maximum value
of $C$ at small time lag, about 55\% is not expected to be 100\%
because the hot-wire signal is not calibrated and its calibration
curve is not expected to be  linear. Furthermore, the noises of the
sensors that have a non-hydrodynamic origin, are not expected to be
correlated and will naturally decrease the correlation coefficient of
the signals.

The most striking result is to obtain nearly the same level of correlation
in He~I and in He~II. This gives a strong indication that the most
energetic velocity fluctuations of the flow are producing similar
response of the hot-wire in He-I -where it is a validated anemometer-
and in He-II.

\section{Discussion}
\subsection{Response without external flow}
\label{par:stat}

Due to the 21 K operating temperature of the hot wire, the 
liquid helium in the vicinity of the wire surface is in the
supercritical normal phase (noted He~I hereafter for simplicity), 
while far from the wire, helium is
in the He-II phase. These near-wire and far-wire regions are separated
by an isotherm interface at  $T_\lambda$.
In the absence of external flow, we will first assume that the
temperature distribution has a cylindrical symmetry, $r_\lambda$
being the radius of the $T_\lambda$ isotherm.

In the cylindrical shell of He-I surrounding the wire, molecular
conduction is responsible for most of the heat transport
\footnote{Comparatively, radiative heat transfer is inefficient due to
  its $T^4$ scaling, as well as natural convection due to the thinness
  of the He-I shell . Indeed, all Grashof numbers assessed are found
  significantly lower than unity, regardless of the chosen fluid
  properties over the interval $T_\lambda - T_w$}. For a typical heat
flux of $\varphi_0=100 W.cm^{-2}$, numerical integration of Fourier
law between $T_w\simeq 21 K$  (at $r=D_w /2$) and $T_\lambda \simeq
2.15 K $  (at $r=r_\lambda$)  gives a He-I shell layer thickness of 
$r_\lambda  - D_w/2\simeq 0.2\,\mathrm{\mu m}$.
Although Fourier law 
may not be accurate over such a small distances and considering the
very large temperature gradient, it still provides an estimate and
shows that the He-I shell surrounding the wire is significantly
thinner than the wire diameter. 

In the far-wire region, molecular conduction is complemented by the
counter-flow heat transport mechanism, specific to He-II
\cite{VanSciverLivre2012}. At scales larger than the
typical inter-vortex spacing, the overall mean heat flux in He-II is
given by :
\begin{equation}
  \label{eq:GM}
  \tilde \varphi =\left(f(T,P)^{-1}\nabla T\right)^{1/m},
\end{equation}
\noindent $m \simeq 3$ is the
so-called Gorter-Mellink exponent (it was shown
  that $m$=3.4 generally leads to better fits of
  experimental data\cite{Sato:AdvCryEng2006, BonMardion79}) 
and $f(T,P)^{-1}$, called the
conduction function, is a highly temperature and pressure dependent
quantity which is null at $T_\lambda$ and maximum at
$1.93\,\mathrm{K}$ in our experimental conditions.

\begin{figure}[ht!]
  \centering
  \includegraphics[width=8.5cm]{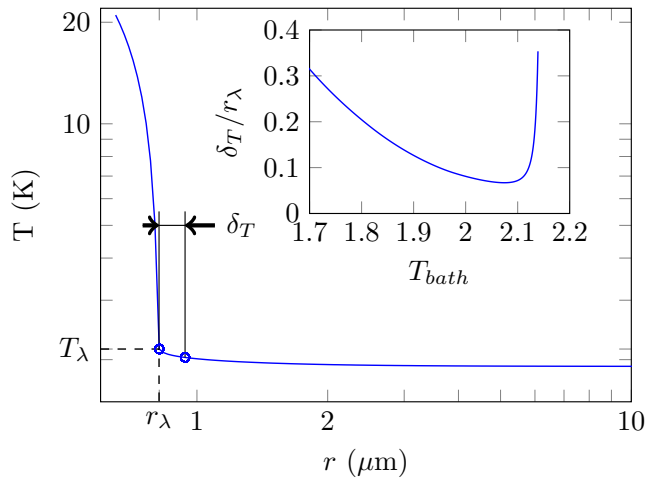}
  \caption{Typical radial temperature profile around the wire, computed for
    $T_{bath} = 1.91$K. Insert: Thermal boundary layer in He~II
    $\delta_{T}$ normalized by $r_\lambda$, the radius of He~I region.}
  \label{fig:thermalBL}
\end{figure}

The efficient heat transport brought by the counter-flow mechanism
(Eq. \ref{eq:GM}) results in significantly smoother temperature
gradients in He-II than in He-I (figure~\ref{fig:thermalBL}). The
insert in figure \ref{fig:thermalBL} illustrates 
the dimensionless thermal boundary layer thickness $\delta_{T} /
r_{\lambda}$ defined as $T(r_\lambda +
\delta_{T})=(T_{bath}+T_\lambda)/2$.  As can be seen, the thermal
boundary layer extends over a typical distance  $\delta_{T} \simeq
r_\lambda /5 \simeq D_w /10$. 

The conduction function is generally used for systems  in which the
heat flux is  lower than in this experiment. In our case the very
high heat flux leads to counter-flow velocities $v_n-v_s =
\tilde \varphi /\rho_sST$ that can be of order of the second sound
velocity near the wire. For this reason, quantities extracted from the
classical literature must be applied to our case with
caution. However, it is interesting to use experimental
(e.g. \citet{Tough82} and references there in) and numerical
(e.g. \citet{Schwarz78}) fits to evaluate the superfluid vortex line
density $\mathcal{L}_0 (r)$ sustained by the counter-flow. Neglecting
the mean velocity of the vortices in the tangle, the vortex line
density can be estimated as: 
\begin{equation}
  \label{eq:vortexdens}
  \mathcal{L}_0 = a \left(v_n-v_s\right)^2
\end{equation}
where $a$ is a temperature dependent parameter. Using numerically
computed values~\cite{Schwarz78} of $a$, the maximum of
$\mathcal{L}_0$ is found at  $r=r_\lambda$, and represents an
inter-vortex spacing of order $r_\lambda /50 \simeq D_w/100$. This
indicates that close to the wire, the inter-vortex spacing is much
lower than the wire diameter. In first approximation,  the time averaged
vortex line density can be approximated as a continuous field at the
wire diameter length scale, which thus justifies \textit{a posteriori}
the use of a continuous model to derive order of magnitude
  estimates.  

Numerical integration of the model in He-I and He-II is done in
cylindrical coordinates using Dirichlet boundary conditions: 
\begin{equation*}
  \left\{ \begin{array}{lcl}
        T=21\,\mathrm{K}& \mbox{for} & r=D_w/2 = 0.65 \,\mathrm{\mu m} \\
      T=T_{bath} & \mbox{for} & r=1000 D_w

\end{array}\right.
\end{equation*}

On figure~\ref{fig:stat}(a) we have represented experimental (line and
circles) and numerical (diamonds) heat flux $\varphi_0$ at the surface
of the wire, as a function of the bath temperature $T_{bath}$. For
temperatures well below $T_\lambda$, we find that our simple model
accounts reasonably well both for the order of magnitude and
temperature dependence of heat flux, showing that the underlying
phenomena are mainly driven by the counter-flow mechanism.
This is consistent with the previous studies with heated
micro-wires\cite{Shiotsu:1994,Ruzhu:1995}, although they were done with wires
with diameters 40 to 60 times larger. 
For bath temperatures $T_{bath}$ close to $T_\lambda$, our model predictions
are underestimated by typically $50\,\mathrm{Wcm^{-2}}$. Since this
offset is also
present right above $T_\lambda$, it is not He-II related and its
origin has not been examined in detail.  Possible origins include
thermal end-effect associated with the prongs or residual offset
introduced by the circuitry. It should also be stressed out that the
accuracy of our numerical simulation depends on the accuracy of the
data we use for the conduction function: for a given 
temperature gradient, the computed flux using 
Hepak\textsuperscript{\textregistered} library (used for the
numerical integration above) and correlations by
\citet{Sato:AdvCryEng2006}  differ by up to a factor of  $\sim 2$ for
large heat fluxes.

As a final remark, the observed heat flux without external flow in
He~II is a  time dependent quantity, as illustrated by the PSD of
figure~\ref{fig:spectra} (filled symbols). Over 2 -3 decades, a $f^{-1}$
power law roughly fits the spectra.
The fluctuating behavior of the heat flux in counter-flows was studied
by various authors (for a review, see \cite{Nemirovskii1995}) and
the cause of fluctuations is understood as turbulent nature of the
counter-flow\footnote{To the best of our knowledge, only spectral
  measurements of 
  the vortex line density have been reported previously.
  In particular, Ref.~\cite{Tough82} reports vortex line density
  fluctuations spectra with a $f^{-1}$ power law followed, after a
  cut-off frequency, by a  $f^{-3}$ power law. In this previous
  experiment, the cut off frequency 
  increases with the heat flux and is thus expected to be very high in
  our experiment where the heat flux is $10^3$ times higher. The
  $f^{-1}$ spectra that we 
  measured is therefore expected to correspond to the  $f^{-1}$ vortex
  line density spectra.}.

\subsection{Response to an external flow}

In the previous section, we have presented three experimental
results that can be interpreted stating that hot-wires can provide a
direct measurement of velocity in He-II. Before discussing the velocity 
dependence, we first discuss (and finally discard) two alternative
interpretations, namely the sensitivity to temperature fluctuations and
to the vortex line density present in the external flow.

As stated above, the temperature sensitivity is of order $\tilde
\gamma\approx -2\mathrm{K^{-1}}$. For the temperature driven
fluctuations to contribute significantly to the measured signal, the
temperature rms fluctuations
$\delta T$ should be of order $\delta T = \left(\varphi (v) -
\varphi_0\right)/(\varphi_0 \tilde \gamma) \approx 30\mathrm{mK}$ at the largest
measured velocities. Such 
large temperature fluctuations are not likely to happen in our flow~:
\begin{itemize}
  \item At large time scales, i.e. for frequencies smaller than 10~Hz,
  the temperature controller maintains the temperature
  within $\pm 0.1\mathrm{mK}$. Furthermore, the
  correlation between the velocity and the temperature at large time
  scale is expected to be null because of the very efficient thermal transfer
  associated with He~II.
  \item For smaller time scales, the local energy dissipation could
  produce temperature fluctuations but their order of magnitude would
  be much smaller. A higher bound for the dissipated energy by unit
  mass may be computed using the pressure loss through the nozzle
  which gives the maximum achievable temperature fluctuations in the
  flow : $\theta'_{max}= V_{nozzle}^2/2Cp \approx 10\mathrm{mK}$,
  where $Cp$ is the constant pressure specific heat of helium. Once
  again the efficient thermalization of the flow due to the counter-flow
  prevents any such high temperature fluctuations at small time/length
  scales.
\end{itemize}
Thus, the temperature effect on the hot wire signal can reasonably be
neglected, at least for the low frequency part of the signal. 

Now the
vortex line density $\mathcal{L}$ of the turbulent external flow could
possibly  contribute to the hot wire signal. Indeed, as mentioned earlier, the
vortex lines are the basic ingredient limiting the heat flow in the
counter-flow surrounding the wire and one could argue that the
time-dependent vortex lines carried by the external flow will add up
to the intrinsic vortex lines generated by the counter-flow. Here
again two main arguments lead us to discard the relevance of this mechanism:
\begin{itemize}
  \item The mean vortex  spacing $\delta = \mathcal{L}^{-1/2}$ was
  found to behave similarly with the Kolmogorov dissipative scale in
  classical turbulence \cite{Salort11,Babuin:EPL2014},
  i.e. $\mathcal{L}$ increases with the Reynolds number as $Re^{3/2}$
  and thus with velocity. This leads to the conclusion that  a higher
  mean velocity would lead to a degraded heat transfer, which is
  obviously opposite to the observation of Fig.~1-b.
  \item Both references cited above provide correlations  between $\delta$
  and the Reynolds number. In our conditions $\delta \approx
  3\,\mathrm{\mu m}$ which is more than 100 times larger that the estimated
  value near the  wire. The vortex line density of the external
  flow is much sparser than the one produced
  by the thermal counter-flow near the wire.
\end{itemize} 

To discuss the velocity dependence, we first come back to the
calibration law Eq.~\ref{eq:calibration}. Obviously the term
$\varphi_0$ accounts for the heat flux at null velocity. 
In classical fluids this term is mainly due to natural convection
whereas in He-II, it is due to the counter-flow mechanism, as seen
previously. 
Contrary to the classical case, this null-velocity term in He-II is
found to remain dominant over the advection term ($\varphi_0 \gg B
v^\alpha$), even in the presence of a vigorous flow slightly exceeding
$1\,\mathrm{ms^{-1}}$. Thus, it is reasonable to assume that the flow
alters only slightly the underlying counterflow heat transport
mechanism in most of the He-II fluid domain. A  self consistency test
of this assumption can be performed by estimating the local Péclet
number defined as the ratio of the advection and counterflow terms 

\begin{equation}
  \label{eq:Pe}
  Pe(r) = \frac{\rho c_p v \left(T(r)-T_{bath} \right)}{\left(f(T,P)^{-1}\nabla T\right)^{1/m}},
\end{equation}
where $\rho c_p$ is the volumetric heat capacity
\cite{VanSciverLivre2012}. Numerical integration shows that $Pe$
remains smaller than one, in agreement with our assumption.

Solving the steady heat equation in He-II at null velocity gives
$T(r)-T_{bath} \sim r^{1-m}$ for $r \gg r_\lambda + \delta_T$  (where
$f(T,P)$ can be taken constant). Due to this pronounced power law
decay ($1-m<-1$), the heat advected by the external flow in the
far-wire region is small compared to the one advected in the wire
vicinity, as can be seen by integration of the numerator of the
$Pe(r)$. Thus, the velocity dependence of the heat transfer must
originate from the near-wire region, say within few $r_\lambda +
\delta_T \simeq 1 \mu m$ from the wire axis. This analysis justifies
the sensitivity to local fluctuations of the velocity, a key property
of classical hot wires. 

Modeling what happens in the micron-size region which surrounds the
wire, and understanding the resulting velocity dependence is delicate and
is beyond the scope of this experimental work. Difficulties arises
because this near-wire region embraces a He-I classical fluid domain
surrounded by the He-II superfluid  and, on top of both the velocity
boundary layer produced by the incoming flow impinging the
wire. Additional technical difficulties arise from the strong fluid
property variations in space, a possible breakdown of the validity of
the continuous model, as well as a spatial cross-over from a radial
counter-flow around the wire to a translational motion of He-II away
from it.

As a final comment, we want to stress the small -if any- temperature
dependence of the advection term $B v^\alpha$ in the hot-wire
calibration (see Fig. 1.b). This experimental result is very
constraining for model development. 
For example, it undermines a straightforward modeling approach
consisting in considering advection as an independent heat transport
mechanism which adds up to an underlying counter-flow
transport. Indeed, simple models elaborated along this line by
integrating the advection term $\rho c_p v \left(T(r)-T_{bath}
\right)$ using a classical velocity boundary layer successfully
predict a $\sim v^{0.5}$ velocity-dependent heat transfer around the
hot-wire but over-estimate significantly the temperature dependence of
the advection term. 

\section{Conclusion}
We have brought three experimental observations, and presented
quantitative arguments, that show  that a hot-wire can be used as a
local velocity sensor in He~II turbulent flows. 

We showed that the calibration law (against the mean velocity) shares
some common scalings  with the one observed in classical fluids. 
The good correlation of the hot-wire signal with a validated local
velocity sensor together with the $f^{-5/3}$ scaling of the power
spectra are further indications that the response of the hot-wire in
He~II is mainly related to the local velocity of the fluid.  

The microscopic mechanisms near the surface of the wire when submitted
to an external flow are delicate to analyze
for a number of reasons listed above. Still, we could justify that the observed
velocity dependence originates in the near vicinity of the wire,
within typically one micron. Thus, the very high effective
conductivity of He-II does not spoil the spatial resolution of the
sensor. 

Although further studies are needed to understand the microscopic
physics at play within the micron-thick shell surrounding the
hot-wire, such probe can already be counted as local velocity sensor
in superfluid helium. 

\subsection{Acknowledgment}
We would like to thank J.~Salort for the design and implementation of
the Pitot tubes used in preliminary studies. We also acknowledge
J.~Duplat and B.~Rousset for fruitful discussions and P.~Charvin for technical
support. This work was supported by the French National Research Agency grant ANR-09-BLAN-0094-01 and by the European Community Framework Programme 7, EuHIT - European High-performance Infrastructures in Turbulence, grant agreement no. 312778.  

\bibliography{HeIIHotwire}

\end{document}